\documentclass[11pt]{article}

\usepackage{amsmath,amssymb,amsthm,fullpage,graphicx}
\usepackage{authblk}
\usepackage{hyperref}
\usepackage{enumitem}

\newcommand{\pref}[2]{\mathsf{pref}_{#1}\!\left(#2\right)}

\title{APoW: Auditable Proof-of-Work Against Block Withholding Attacks}

\author[1,2]{Sergio Demian Lerner\thanks{\href{mailto:sergio@fairgate.io}{sergio@fairgate.io}}\hspace{3pt}} 

\affil[1]{Fairgate Labs} 
\affil[2]{Rootstock Labs}
\date{}

\newcommand{\APoW}{APoW }
\newcommand{\keywords}[1]{%
  \vspace{0.5em}
  \noindent\textbf{Keywords:} #1
}

\begin{document}

\maketitle

\begin{abstract}
We introduce Auditable Proof-of-Work (APoW), a novel proof-of-work (PoW) construction inspired by Hashcash-style nonce searching, which enables the auditing of other miners’ work through accountable re-scanning of the nonce space. The proposed scheme allows a miner to probabilistically attest to having searched specified regions of the nonce space in earlier mining rounds, while concurrently earning rewards for performing productive work for a new block or pool share. This capability enables miners belonging to a mining pools to audit another miner’s claimed effort retroactively, thereby allowing the probabilistic detection of block withholding attacks (BWAs) without requiring trusted hardware or trusted third parties. As a consequence, the construction supports the design of decentralized mining pools in which work attribution is verifiable and withholding incentives are substantially reduced. The scheme preserves the fundamental properties of conventional PoW, including public verifiability and difficulty adjustment, while adding an orthogonal auditability layer tailored to pool-based mining. Finally, while a full deployment of \APoW in Bitcoin would require a consensus rule change and minor modifications to mining Application-Specific Integrated Circuits (ASICs), the construction remains practically useful even without consensus changes, for instance, as a pool-level auditing mechanism that enables verifiable pay-for-auditing using existing pool reserves.

\end{abstract}

\keywords{Proof-of-Work, Bitcoin, Auditable Proof-of-Work; APoW; Block
Withholding Attacks; Mining Pools; Decentralized Mining; Hashcash;
Nonce Searching}

\tableofcontents

\section{Introduction}

Mining pools\cite{palatinus2010miningpool,rosenfeld2011pool} are the dominant coordination mechanism in proof-of-work systems such as Bitcoin\cite{nakamoto2008bitcoin}, allowing individual miners to reduce variance by aggregating hashpower and sharing rewards proportionally to contributed work. In a typical pool protocol, workers repeatedly attempt proof-of-work puzzles and submit shares—partial solutions that meet a reduced difficulty threshold—to demonstrate ongoing effort. When a full solution is found, the corresponding block is broadcast to the network and the resulting reward is distributed among the participants according to the pool’s payout policy.

A Block Withholding Attack\cite{rosenfeld2011pool,eyal2014minersdilemma} (BWA) arises when a malicious worker (or coalition of workers) deliberately suppresses valid full-difficulty blocks while continuing to submit low-difficulty shares to the pool. From the pool’s perspective, the attacker appears to contribute honestly: shares are valid, timely and statistically consistent with the expected effort. However, by withholding blocks, the attacker reduces the effective block discovery rate of the pool, thereby harming other participants or enabling strategic advantages in competitive settings.

BWAs are particularly difficult to detect because, under standard proof-of-work constructions, the absence of a block is not itself evidence of misbehavior. Block discovery is a low-probability event, and variance alone can explain extended periods without successful blocks even for honest miners. Consequently, distinguishing deliberate withholding from statistical fluctuation requires long observation windows and provides, in the best case, weak confidence guarantees. As a result, most existing pool protocols cannot attribute blame or provide cryptographic evidence of withholding behavior.

The problem is exacerbated in decentralized or trust-minimized pool designs, where pool operators lack both identity-based enforcement mechanisms and trusted execution environments. Traditional countermeasures—such as reputation systems, payment delays, or centralized monitoring—either reintroduce trust assumptions or remain vulnerable to Sybil attacks and strategic churn. 

Fundamentally, BWAs exploit a structural limitation of conventional proof-of-work schemes: while valid solutions are publicly verifiable, unsuccessful search effort leaves no auditable trace. 

While zero-knowledge proofs could, in principle, certify exhaustive search without revealing individual attempts, constructing such proofs for hash-based PoW entails prohibitive computational and protocol overhead. Pools must therefore infer honesty indirectly from share statistics, an approach that is inherently probabilistic and slow to react. This limitation motivates the exploration of proof-of-work constructions that preserve standard validation properties while enabling miners to later demonstrate, in a verifiable manner, that specific regions of the nonce space were genuinely explored.

In this work, we address this gap by introducing a proof-of-work mechanism that supports auditable reuse of prior nonce-space, allowing pools or third parties to probabilistically verify past mining effort while new work is being performed. This added auditability enables the detection of block withholding behavior without requiring trusted hardware, interactive protocols, or changes to the network’s block validation rules, thereby opening the door to decentralized mining pools with stronger resistance to BWAs.

\subsection{Mining Pool Payment Schemes and Exposure to Block-Withholding Attacks}
\label{sec:prior-art-pool-schemes}

Mining pools employ a variety of reward distribution schemes to compensate participating miners for contributed hash power. These schemes differ significantly in how mining variance is allocated between miners and pool operators, and consequently in their susceptibility to block-withholding attacks (BWAs), in which a miner submits valid low-difficulty shares while deliberately withholding full-difficulty blocks.

\paragraph{Block-withholding incentives.}
A key observation from prior work is that the profitability of a BWA depends critically on whether a miner's payout is conditioned on the pool's actual block discovery events. If rewards are independent of block discovery, a withholding miner can preserve their expected income while reducing the pool’s revenue, thereby creating a strictly profitable deviation.

\paragraph{Classification of payment schemes.}
We classify the most widely deployed mining pool payment schemes according to their resistance to BWAs:

\begin{itemize}
    \item \emph{Block-conditioned schemes}, such as Pay-Per-Last-$N$-Shares (PPLNS) and score-based variants, distribute rewards only when the pool successfully mines a block. In these schemes, withholding a valid block reduces the attacker’s own expected revenue, rendering BWAs economically irrational.
    \item \emph{Share-conditioned schemes}, such as Pay-Per-Share (PPS), compensate miners solely based on submitted shares, regardless of block discovery outcomes. These schemes transfer all variance risk to the pool operator and make BWAs strictly profitable.
    \item \emph{Hybrid schemes}, such as PPS+, partially condition rewards on block discovery (typically for transaction fees), resulting in intermediate exposure to BWAs.
    \item \emph{Full Pay-Per-Share (FPPS)} schemes extend PPS to both the block subsidy and transaction fees by paying miners the expected value of each per share, thereby maximizing pool-side variance and BWA exposure.
\end{itemize}

\paragraph{Security comparison.}
Table~\ref{tab:bwa-resistance} summarizes this classification.

\begin{table}[h]
\centering
\begin{tabular}{lccc}
\hline
\textbf{Scheme} & \textbf{Variance Bearer} & \textbf{BWA Resistance} & \textbf{BWA Profitability} \\
\hline
PPLNS & Miner & Strong & No \\
Score-based & Miner & Strong & No \\
PPS+ & Shared & Moderate & Partial \\
PPS & Pool & Weak & Yes \\
FPPS & Pool & Very weak & Yes \\
\hline
\end{tabular}
\caption{Classification of mining pool payment schemes by resistance to block-withholding attacks.}
\label{tab:bwa-resistance}
\end{table}

\paragraph{Empirical exposure of Bitcoin hashrate.}
Based on public disclosures by major mining pools, historical hashrate distribution data, and advertised payout options, we estimate that approximately \textbf{25\%--35\% of the current Bitcoin network hashrate} participates in pools offering PPS or FPPS-style payouts, and is therefore \emph{economically susceptible} to block-withholding attacks in the absence of additional detection or auditing mechanisms.
This estimate accounts for large pools that either operate exclusively under PPS/FPPS or offer these schemes as a selectable option, and excludes pools that rely solely on PPLNS or score-based methods.

While the exact fraction varies over time with market conditions and miner preferences, the prevalence of PPS-style payouts highlights a fundamental tension between payout predictability and incentive compatibility in mining pool design.

\paragraph{Implications.}
The above classification illustrates that widely deployed pool payment mechanisms remain vulnerable to BWAs unless miner compensation is explicitly tied to verifiable block discovery. This motivates the exploration of alternative constructions, such as auditable proof-of-work mechanisms, that can preserve predictable payouts while restoring incentive compatibility without requiring trusted pool operators or trusted hardware.

\section{Prior Work on Deterring BWA}
\label{sec:prior-work-bwa}

Block withholding attacks (BWAs) exploit a structural feature of pooled mining: a worker can
appear productive by submitting low-difficulty shares while selectively suppressing full-difficulty
block solutions, thereby reducing the pool's block revenue without immediately revealing
misbehavior. Since block discovery is a rare event, a pool observing ``too few'' blocks cannot,
in general, distinguish malice from variance using only standard share statistics.

\subsection{Rosenfeld's Oblivious Shares Proposal}
\label{sec:rosenfeld-oblivious}

Rosenfeld's early analysis of pooled mining identifies block withholding as a practical deviation
available to dishonest pool participants and surveys several mitigation ideas
\cite{rosenfeld2011pool}. In \S6.2.3, Rosenfeld proposes \emph{oblivious shares}, a protocol-level
mechanism intended to make it possible for a pool to determine whether a submitted share is also
a valid block, while preventing the miner from knowing this fact at the time of discovery.
Intuitively, the pool supplies work that is constructed so that only the pool can recognize
full-block solutions among the shares it receives; thus, a miner cannot selectively withhold blocks
because it cannot identify them.

\paragraph{Limitations and deployment frictions.}
While conceptually appealing, oblivious shares introduce several important constraints that are
in tension with modern pooled-mining practice:

\begin{enumerate}
  \item \textbf{Hard fork / header reinterpretation.}
  To apply an oblivious-shares-style mechanism to  Bitcoin, a hard-fork is  required: a difficult-to-achieve network-wide coordination procedure. Both full nodes and SPV/light
  clients would need to upgrade to preserve their security model, since the proof-of-work
  predicate they implicitly rely on is tied to header semantics. Alternative variants that avoid
  requiring light-client upgrades have been discussed, but at a cost of significantly reduced
  security for those clients \cite{optech315,towns2024bitcoindev}.

  \item \textbf{Share-submission latency versus immediate block broadcast.}
  Under Stratum~v2 in certain modes, a worker that finds a full block can broadcast it directly to
  the P2P network so propagation begins immediately, potentially before the share reaches the pool
  server. With oblivious shares, by design, the worker cannot recognize a block solution; the share
  must first be delivered to the pool and transformed into a full block before it can be broadcast,
  introducing additional latency in the best-case honest setting \cite{optech315}.

  \item \textbf{Pool centralization.}
    Rosenfeld’s oblivious-shares approach relies on the pool having secret information that allows it to distinguish block-valid shares from ordinary shares. This requirement naturally concentrates authority at the pool operator, who must generate private templates, control share interpretation, and act as the sole entity capable of recognizing and broadcasting valid blocks. As a result, the protocol structure inherently favors centralized pool architectures. 

\end{enumerate}

\subsection{Rosenfeld's “pop quiz” Proposal}

\label{sec:rosenfeld-popquiz}

In addition to oblivious shares, Rosenfeld discusses an alternative deterrence mechanism informally referred to as a \emph{``pop quiz''} approach \cite{rosenfeld2011pool}. The idea is to occasionally challenge miners with tasks intended to test whether they are behaving honestly, requesting them to scan a nonce space for which the pool already knows contains a solution.  

This method has an important limitation: quiz challenges are often distinguishable from ordinary mining work. In practice, a miner receiving a quiz template can detect that it does not correspond to the current tip of the chain (e.g., due to a stale block height or an unexpected parent hash). Once a miner recognizes that a task is a quiz, he may behave honestly, avoiding detection.

\subsection{Peer Reviews and Short-Lived Audits Proposal}

An idea similar to the pop quiz is to temporarily perform "peer reviewed" mining: the pool requests that two competing miners mine the same nonce range for a short period. We can also use the mechanism as an audit scheme, like ours, by requesting a miner to re-scan the nonce space range scanned by another miner in a previous block.

However, the pool cannot dedicate much resources to pop-quizzes or audits without substantial losses, and the probability of catching a cheater is very low. Consequently, while pop quizzes and audits may provide a weak probabilistic deterrent, they usually cannot collect cheating evidence before the pool goes bankrupt.

Additionally, audit/review hashing does not contribute to the effective hash rate of the pool, so the miner can rationally choose to ignore the challenge entirely.

\section{Description}

\APoW permits miners to operate in two distinct modes. In \emph{normal mining mode}, a miner searches for a proof-of-work solution for the current block in a manner analogous to Bitcoin. In \emph{audit mining mode}, a miner simultaneously performs productive mining and audits prior mining effort, an activity we refer to as \emph{verification mining} or \emph{v-mining}. A miner engaged in v-mining is called a \emph{v-miner}, and a block produced as a result of v-mining is termed a \emph{verification block} (v-block).

A miner may enter or exit v-mining for variable durations, depending on pool policy and auditing requirements. We define a \emph{mining round} $i$ as the mining activity associated with block height $i$. For a miner $A$, the \emph{search space} $S_i^A$ in round $i$ is defined as the set of nonce intervals explored by $A$ during that round. 

Each element of $S_i^A$ is called an \emph{auditable work unit} and is represented by a tuple $(G, n_s, n_e)$, where $G$ is a block template header, $[n_s, n_e]$ is a contiguous nonce interval scanned under that template, and all templates $G$ appearing in auditing work units of $S_i^A$ share the same parent block.

For clarity of exposition, we assume a single nonce field of sufficient precision, even though Bitcoin conceptually splits the nonce into an \emph{extraNonce} (most significant bits) and a \emph{nonce} (least significant bits). This abstraction does not affect the generality of the construction.

When miner $A$’s work in round $i$ is audited by another miner $B$, the auditor does not re-execute the entire search space $S_i^A$. Instead, $B$ audits only a randomly selected subset of $S_i^A$, yielding a probabilistic verification of $A$’s claimed mining effort.

In Section~\ref{sec:nonce-assignment}, we describe how the search space $S_i^A$ associated with a miner $A$ is communicated to the mining pool and subsequently assigned to an auditor in a centralized pool setting. We further discuss how analogous information can be disseminated in a decentralized mining pool, allowing multiple participants to independently reason about and audit the claimed search space $S_i^A$.

A v-miner $B$ can v-mine in round $j$ ($0 < j - i \leq D$), where $D$ is the maximum depth, by reusing a block template header $G$ drawn from a tuple in $S_i^A$. During v-mining, miner $B$ searches for hash outputs satisfying two predicates simultaneously. The first predicate requires the hash digest to match an unpredictable bit pattern $b_2$ of length $d$; satisfying this predicate yields ordinary shares or full block solutions that are credited to the v-miner under the pool’s standard accounting rules. The second predicate requires the hash digest to match a distinct pattern $b_1$ consisting of $d$ zero bits; this predicate is used exclusively for auditing purposes.

Hash outputs satisfying the auditing predicate are reported to the mining pool as evidence of v-mining progress but do not contribute to block discovery or pool rewards beyond their role in verification. If the pool observes that the v-miner discovers auditing shares or full solutions within a nonce interval previously claimed as explored by miner $A$, and these findings were not reported by $A$ during round $i$, the pool obtains cryptographic evidence of withholding behavior and may apply penalties according to its payout policy.

Because the v-miner reuses a block template header $G$, the v-miner cannot commit to the new block template in the same way as a normal miner. We can add a method to associate the block with extra data post-PoW or to disallow extra data. This results in two scheme designs:

\begin{itemize}
    \item \textbf{Association with extra data.} The v-miner proceeds to \textit{seal} a block: he signs the block template with a public-key scheme and broadcasts the block signature along the proof-of-work and the signed block template. The v-block can be non-empty.

    \item  \textbf{No extra data.} A v-block must not carry neither transaction nor block time information. The v-block is empty.
\end{itemize}

\subsection{Scheme 1: Non-Empty V-Blocks}

Let us assume that we associate a new block header with the PoW after it is found. Then the parent block will also be chosen post-PoW, which means that a v-block could be sealed to fit in different slots of the blockchain. This could disrupt the incentives of mining. A careful analysis is required including how this impacts mining fairness and selfish mining attacks. Without signature equivocation and penalization of a stake, generating valid v-blocks is free.  To prevent attacks, we add a blockchain consensus rule that a v-block can be included only just immediately after its audited block. Without this consensus rule, the protocol would need to check each new v-block against all past v-blocks to prevent double-inclusion. The downside of the immediate inclusion rule is that the audit time is reduced to a single block interval, which can vary widely. 

Although the v-miner could seal multiple competing blocks for the same block height for free, this attack can only result in short-lived forks, and the network converges at the following normal block. By disallowing consecutive v-blocks, we avoid longer forks.  

\subsection{Scheme 2: Empty V-Blocks}

When a v-block does not carry neither transaction nor block timestamp information, the v-block does not need sealing and is already unique. Although this option is possible, it has the downside that it reduces the blockchain transaction throughput. We analyze this variant in Appendix \ref{app:unsealed-vblocks}.

We now continue the description of Scheme 1.

\subsection{Fairness}

For v-block mining to be fair, a miner should not be able to create multiple valid hash digest targets. That would increase the success probability for each nonce. In particular, a miner should not be able to solve multiple consecutive blocks in parallel with the same nonce scan

\begin{figure}[h]
\centering
\includegraphics[width=0.8\textwidth]{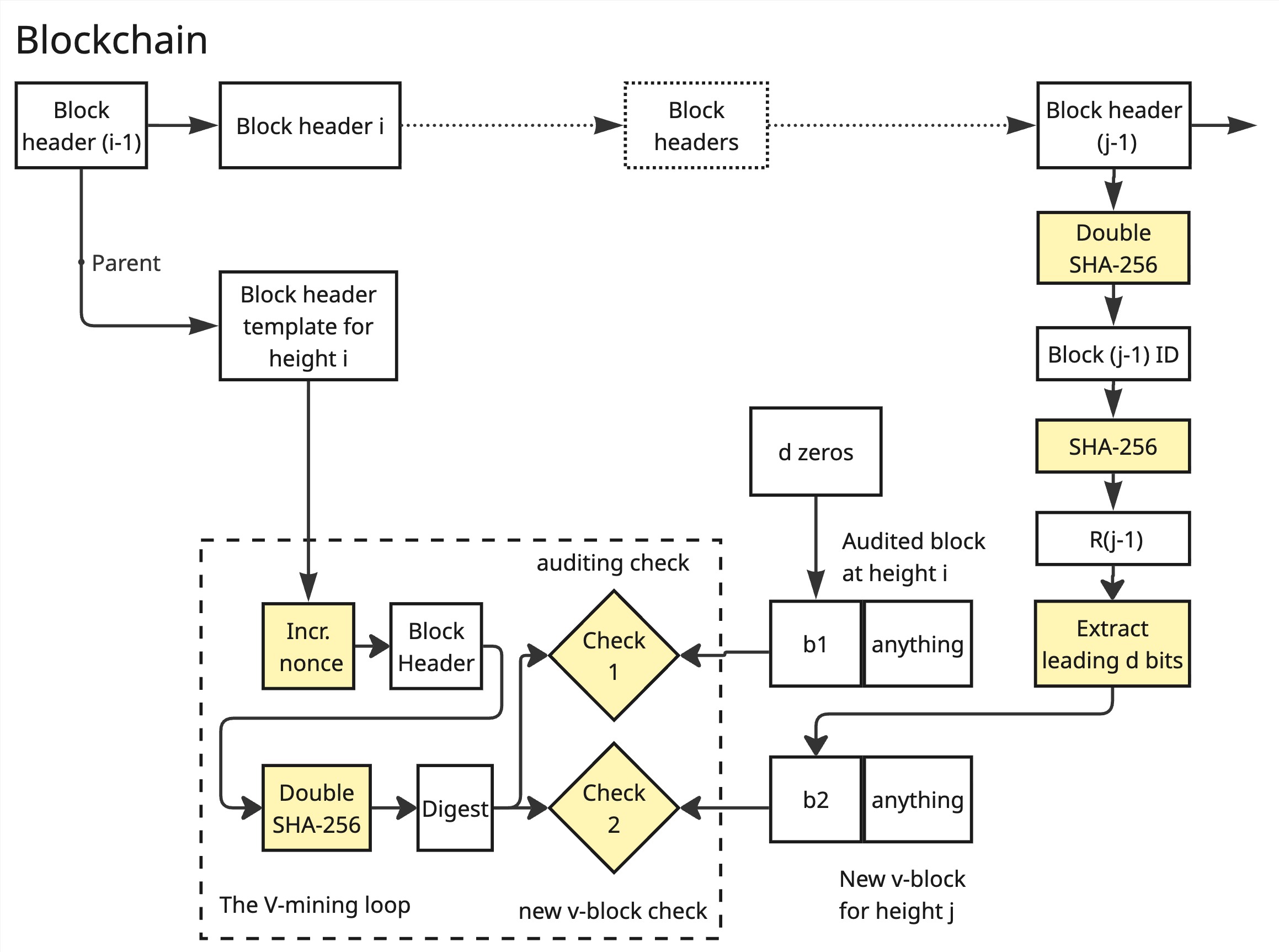}
\caption{The V-Mining Process Adapted for Bitcoin (simplified)}
\label{fig:V-Mining}
\end{figure}

Let us assume that the audit is conducted while mining a block at height $j$ ($0 < j - i \leq D$), and the miner round audited is at height $i$, then we choose $b_2$ as the leading $d$ bits of $H(B_{ID}(j-1))$, where $B_{ID}(j-1)$ is the block ID of the block on the blockchain at height $j-1$. This dependency prevents the miner from obtaining the pattern $b_2$ before a block at height $b_2$ has been found. The  $b_1$ will consist of $d$ zeros. Figure ~\ref{fig:V-Mining} illustrates the v-mining process using Bitcoin PoW function.

Fixing the pattern $b_2$ prevents a malicious miner from generating tables of valid hash digest by grinding the block templates, and then performing several parallel checks for each nonce iterated.

Because the proof-of-work cannot be directly linked to the identity of the v-miner, an explicit binding mechanism is required for the consensus protocol to issue rewards correctly. Without such a mechanism, an external observer could extract a v-block header from the mempool and resubmit it—potentially with a higher transaction fee—in order to claim the reward.

In the following sections, we will describe two of these mechanisms: one based on the pre-commitment of the auditor payout address and another based on zero-knowledge proofs.

\subsection{Notation}
Let
\[
H:\{0,1\}^\star \rightarrow \{0,1\}^\lambda
\]
be a cryptographic hash function with output length $\lambda$.

Let $i \in \mathbb{N}$ denote the block height.
The block template header at height $i$ is denoted by
\[
G_i := ( \ \mathsf{version}_i,\  \mathsf{parent}_i, \ \mathsf{txroot}_i,\ \mathsf{height}_i,\ \mathsf{time}_i,\ \mathsf{difficulty}_i, \mathsf{poolAddress}_i)
\]
The field $\mathsf{height}_i$ is equal to $i$. Note that the nonce is not part of the template header. 

For a bitstring $x$, let $\pref{k}{x}$ denote its first $k$ bits

We define two algorithms: mining and v-mining.

\subsection{Mining Algorithm}

Let $n$ be a nonce of a predefined length $l_n$, and let $i$ be the block height. Let the message $M$ be
\[
M := G_i \,\|\,  n .
\]
The value $G_i$ bounds the message $M$ to a particular block height. In the case of Bitcoin, the block height is embedded in the coinbase transaction. 
 
The miner samples $n \leftarrow \{0,1\}^{l_n}$ and computes
\[
X := H(M)
   = H(G_i \,\|\,  n )
\]
until the work condition 
\begin{equation}
\label{eq:stage3}
 X =  z \,\|\, y_m 
\end{equation}
is satisfied, where $z$ is a zero bitstring, $z \in \{0,1\}^{d}$ and $y_m \in \{0,1\}^{\lambda - {d}}$. In Figure ~\ref{fig:V-Mining}, $b_1 = z$.

\subsection{V-Mining Algorithm}

A v-miner auditing a tuple ($G,n_s,n_e$) of $S_{i}^A$ at height $i$, while mining a v-block for height $j$ ($ j > i)$. The v-miner derives the bit strings $b_2$ from $B_{ID}(j-1)$, the block ID on the canonical blockchain at height $j-1$.

We define $b_2$ as

\begin{equation}
\label{eq:stage1}
 b_2 = \pref{d}{H(B_{ID}(j-1)) }
\end{equation}

\paragraph{Proof-of-work search.}

For each nonce test, we build the message $M$ from $G$ and the nonce. The v-miner samples $n \leftarrow \{0,1\}^{l_n}$ in the range $[n_s,n_e]$ and computes
\[
X := H(M)
\]
as normal, until the work condition 
\begin{equation}
\label{eq:stage3}
 X =  b_2 \,\|\, y_m 
\end{equation}
is satisfied, with any bit-string $y_m \in \{0,1\}^{\lambda - {d}}$.

To provide a finer-grained measurement of difficulty, instead of difficulties that can be only powers of 2, we can compute a maximum distance of the target from the $d$-bit truncated hash digest to the target value $u$. While the powers-of-2 difficulty adjustment method was used in pre-release versions of Bitcoin, the launched version of Bitcoin uses the distance from the hash digest to zero.

The pattern matching method can be used by the mining hardware only for some leading bits of $b_2$, eliminating the need for a more expensive subtraction operation in the ASIC.  

\paragraph{Sealing and validation.}

After finding the PoW, the v-miner builds v-block which consists of the tuple ($i, G_i,n, j, G_{j}, \sigma$) where $G_{i}$ is the template audited at height $i$, $G_{j}$ is the new block template for height $j$ and $\sigma$ is a signature of the tuple (excluding the signature field), for the public key associated with the auditing of $G_A$. The nonce $n$ is valid if it results in the expected PoW for template $G_A$ with the target pattern $b_2$, according to the difficulty for a block at height $i$. The field $G_{i}$ is valid if it satisfies all the conditions of a block header for height $i$ except for its modified PoW. For $G_i$ the $\mathsf{poolAddress}$ must be a valid payment address for the ledger. This is the address used to receive the v-block reward.

The field $G_{j}$ is valid if it satisfies all the conditions of a normal block for height $j$. No PoW is required for this template (the PoW was delegated to $G_{i}$). The $\mathsf{poolAddress}$ field of $G_{j}$ is ignored.

\subsection{Avoidance of Long Forks}

In the absence of explicit penalties for equivocation, a miner may attempt to submit multiple competing v-blocks at the same height~$j$, differing for instance in their transaction sets or timestamps. Short-lived forks of depth~1 are not inherently problematic and are common in Nakamoto-style consensus.

Several hybrid consensus protocols that combine proof-of-work and proof-of-stake, such as 2-hop~\cite{hop2} and TwinsCoin~\cite{twins}, explicitly tolerate single-block forks by enforcing that the block following
a PoS-extended block must be a standard PoW block. This structural constraint prevents the
formation of long runs of stake-extended blocks and ensures that forks cannot grow unboundedly
without additional PoW investment.

\begin{figure}[h]
\centering
\includegraphics[width=0.8\textwidth]{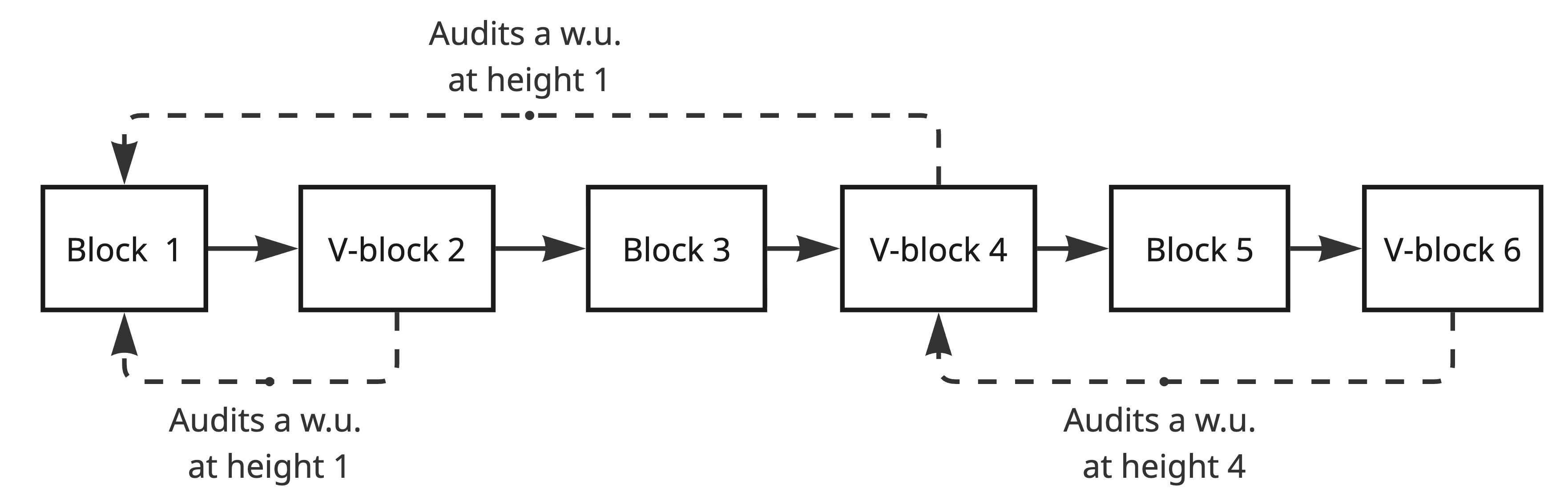}
\caption{A blockchain mixing blocks and v-blocks auditing previous work units (w.u.), with v-blocks immediately followed by a regular PoW block.}
\label{fig:blockchain}
\end{figure}

We adopt a similar rule: any v-block must be immediately followed by a regular PoW block. As a consequence, not all auditable work units associated with a v-block can necessarily be audited in the immediately following round. However, these units remain eligible for auditing in subsequent rounds. Figure~\ref{fig:blockchain} shows a valid blockchain mixing block and v-blocks without consecutive v-blocks.

More generally, we allow auditing to be performed up to $D$ rounds after the corresponding block is mined. This bounded auditing window preserves chain growth while preventing the accumulation of long forks induced by equivocation at the v-block layer.

\subsection{V-Block Attribution}

We present three scheme variants to attribute a v-block and be able to pay its reward. 

\paragraph{Paying to the pool.} Every block $i$ contains the field $\mathsf{poolAddress}_i$ in its block template. This is the address where v-blocks are paid. The address is extracted from the message $M$ in the v-block. The address can be controlled by the pool server, or it can be a smart contract controlled address in the mainchain (if it has this capability) or it can be a sidechain exit address that forwards the funds to a smart contract on a sidechain. The pool server or pool smart contract is responsible for sharing the reward among the workers, in the same manner as a normal block reward. The server can also track who was the auditor of the audited block and pay a special prize to that worker. Alternatively, a field $\mathsf{lastBlock}_i$ could be added to the block template. This field would indicate the block ID of the last block created by the same pool that was included in the canonical chain. The protocol could retrieve that block coinbase transaction and distribute the reward to the same addresses and in the same proportion used in the referenced block. This enables Bitcoin decentralized mining pools that pay their workers directly from the coinbase transaction using multiple outputs.

\paragraph{Paying an auditor whose identity is derived from proof-of-work.}

The scheme can embed auditor selection directly into the consensus protocol, allowing the identity of the auditor to be deterministically known and derived from the blockchain itself. To this end, we extend the block template with an additional field $\mathsf{auditorsRoot}$, defined as the Merkle root of a set of public keys (or ledger addresses) corresponding to all eligible auditors within a mining pool. This field must be built by the mining pool itself and forwarded to all miners.

When auditing a share at height $i$, the identifier of block $i$ already recorded on the blockchain is used as a seed to a cryptographically secure pseudorandom number generator (CSPRNG) that selects a single auditor from the committed set. As a result, immediately after block $i$ is published, the pool can determine which miner is eligible to act as auditor during the mining of block $i+1$.

If a v-block is subsequently produced, it must include a Merkle inclusion proof linking the selected auditor’s public key to $\mathsf{auditorsRoot}$, thereby allowing the consensus protocol to verify both the correctness of the auditor selection and the legitimacy of the reward recipient. The v-block can also be sealed with the auditor's private key, committing to a new block template for height $i+1$ to be processed with the v-block.

\paragraph{Paying to any auditor.} The v-miner can include in the transaction $T$ a zero-knowledge proof that he knows $M$, while revealing all its components except for the nonce. The v-miner can associate the proof with his public address, by proving its knowledge and showing it. Nobody can take attribution of the v-block without finding the nonce themselves. The $\mathsf{poolAddress}_i$ field is not required in this case. A generalized formalization of a ZK-PoW is presented in the Appendix \ref{app:znp}.

\subsection{Pool Share Difficulty Management}

The difficulty of mining a block is $2^d$  (expressed in base-1 difficulty). By requesting the matching of a subset of bits in $b$ instead of all of them, the mining pool can control the difficulty of a share coming from a v-miner.

\section{\APoW Compared to Prior Work}

We compare our scheme with the existent solutions to BWA.

\subsection{Oblivious Shares vs Our Approach}
\label{sec:compare-ours}

Rosenfeld's oblivious shares scheme aims to hide from the miner whether a given share is a valid
block, thereby preventing selective suppression. In contrast, our construction targets a different
axis: it enables auditing of claimed historical search effort by allowing miners to
probabilistically attest, while mining current work, to having previously scanned specified regions
of nonce space for earlier blocks or shares.

This distinction matters operationally. Oblivious shares require the pool to maintain secret
information (and typically to control template construction) so that the miner cannot classify
solutions at discovery time. Our design instead provides an auditability layer that can be
applied retroactively: miners can be challenged to produce evidence consistent with prior scanning while still producing valid shares that will be rewarded by the pool. As a result, our construction supports decentralized pool designs in which miners retain autonomy and no single party must be trusted to correctly identify or disclose block solutions.

While our approach avoids several sources of centralization inherent in oblivious-share constructions, it is not without significant deployment friction for existing blockchains. In particular, upgrading Bitcoin to use the proposed proof-of-work function constitutes a consensus-level change and, therefore, requires a hard fork of the Bitcoin consensus protocol. As a consequence, both full nodes and light clients must upgrade to correctly interpret and validate the new proof-of-work predicate. Moreover, because the construction departs from existing SHA-256–based mining by replacing zero bit checks with bit pattern matches, it may necessitate the deployment of new specialized hardware or ASIC designs adapted to the revised workload. These requirements imply a higher coordination and adoption threshold than approaches that operate purely at the mining-pool protocol layer. Consequently, any practical deployment on top of an existing blockchain would need to demonstrate that the benefits of block withholding deterrence and decentralized pool support outweigh the substantial costs associated with network-wide consensus changes and hardware transition.

\subsection{“Pop quiz” (and its variants) vs Our Approach}
\label{sec:compare-popquiz}

Rosenfeld’s pop-quiz proposal and our construction share a common intuition: \emph{randomized auditing} can be used to deter block withholding even when misbehavior cannot be cryptographically detected. The key difference lies in how the verification cost is handled.

In pop-quiz schemes,  the hashing spent verifying other miners is not rewarded by the blockchain, limiting their practical effectiveness as a BWA deterrent.

In contrast, our approach enables miners to reuse prior nonce-space exploration while mining new blocks or shares. Importantly, this verification occurs \emph{concurrently} with productive mining, rather than as a separate re-mining step, and therefore does not negate the economic efficiency of pooled mining.

\section{Security}

We identify three potential attacks related to the fairness of mining and v-mining, and we show how our scheme resist such attacks.

\paragraph{PoW results caching.} A miner could try to store all $X$'s computed during the mining of a previous block, and when mining a new block, she could fetch the $X$'s from memory, instead of hashing. Using state-of-the-art technology, memory access is at least an order of magnitude slower than computing a hash operation in a pipelined ASIC. Accessing RAM also consumes between one and two orders of magnitude more electricity than computing the hash on-the-fly. Also, memory has to be accessed twice: one for storing and another for loading. Nevertheless, RAM cost would still make this mining method astronomically expensive. An advanced Bitcoin miner performs ~200 TH/s and must be run for 600 seconds on average until a new block is found. Assuming 256 bits per hash digest, this implies storing ~4 exabytes (~$2^{62}$ bytes). That amount of RAM would make a v-mining machine cost more than 10 Trillion dollars as of 2025.  

Memory-CPU trade-off provides little help with regard to RAM costs. Assuming that all hash operations are stored sequentially, each cached hash operation $j$ requires the storage of $X_j$, which is 256-bits in length. It is neither necessary nor optimal to store $X_j$ in full. The digests $X_j$ could be partially stored and recomputed in full after a positive match with the partial target pattern, to verify the full match. For example, if we store 16-bits per digest, we can retrieve 8 hash results per RAM memory access (assuming a 128-bit wide data bus). This results in an 8X speedup for RAM access and still leaves plenty of time for a CPU or a non-pipelined hashing core to recompute promising partial matches. Since the nonce candidates to check will not be sequential, they need to be transmitted to the hashing component, thus the communication can become the system's bottleneck. While this optimization could make storing and retrieving results as fast as hashing, the v-mining machine would still cost over 1T dollars. 

The attacker can also store just 1 bit per hash, packing bits in a 128-bit hint word that can be retrieved in a single RAM access. The attack requires the building of a special ASIC architecture that reads hint words, scans its bits while it increments the nonce counter at double clocking speeds. This small overclocked circuit would stall for 1 cycle when a valid bit (0 or 1) is found in the hint bit sequence, so that the corresponding nonce enters the hashing pipeline, keeping it almost full. Although this design may work, the machine would still cost 39B USD due to its large RAM requirement.

\paragraph{BWA while v-mining.} A malicious v-miner could withhold solutions while v-mining. Since normally the role of auditor would be rotating among all miners, and the hashrate directed towards auditing would be only a small portion of the total hashrate, the impact of a BWA during auditing may be negligible. Nevertheless, we can also change the scheme to enable v-miners to be audited. Let us assume that the block at height $i$ was audited at height $k$ and that auditing process is audited at height $j$ with ($i < k < j$). These are the changes required: 

\begin{itemize}
\item The new consensus rule allows v-blocks auditing other v-blocks to be accepted and equally rewarded, up to a maximum audit block depth $D_2$ ($0 < j - k \leq D_2$). To derive a pattern $b_2$ for producing new blocks at height $j$, we hash the blockchain block at height $j-1$. 

\item The ASIC comparison circuit is modified to allow the checking of two leading bit patterns ($b_1$ and $b_2$). Only when a v-miner $B$ audits a normal miner, the pattern $b_1$ is zero. When a v-miner $B$ audits a v-miner $A$, the pattern $b_1$ will match the pattern $b^A_2$ used by miner $A$ to find a block at height $k$ (derived from blockchain block $k-1$).

\end{itemize}

\begin{figure}[h]
\centering
\includegraphics[width=0.8\textwidth]{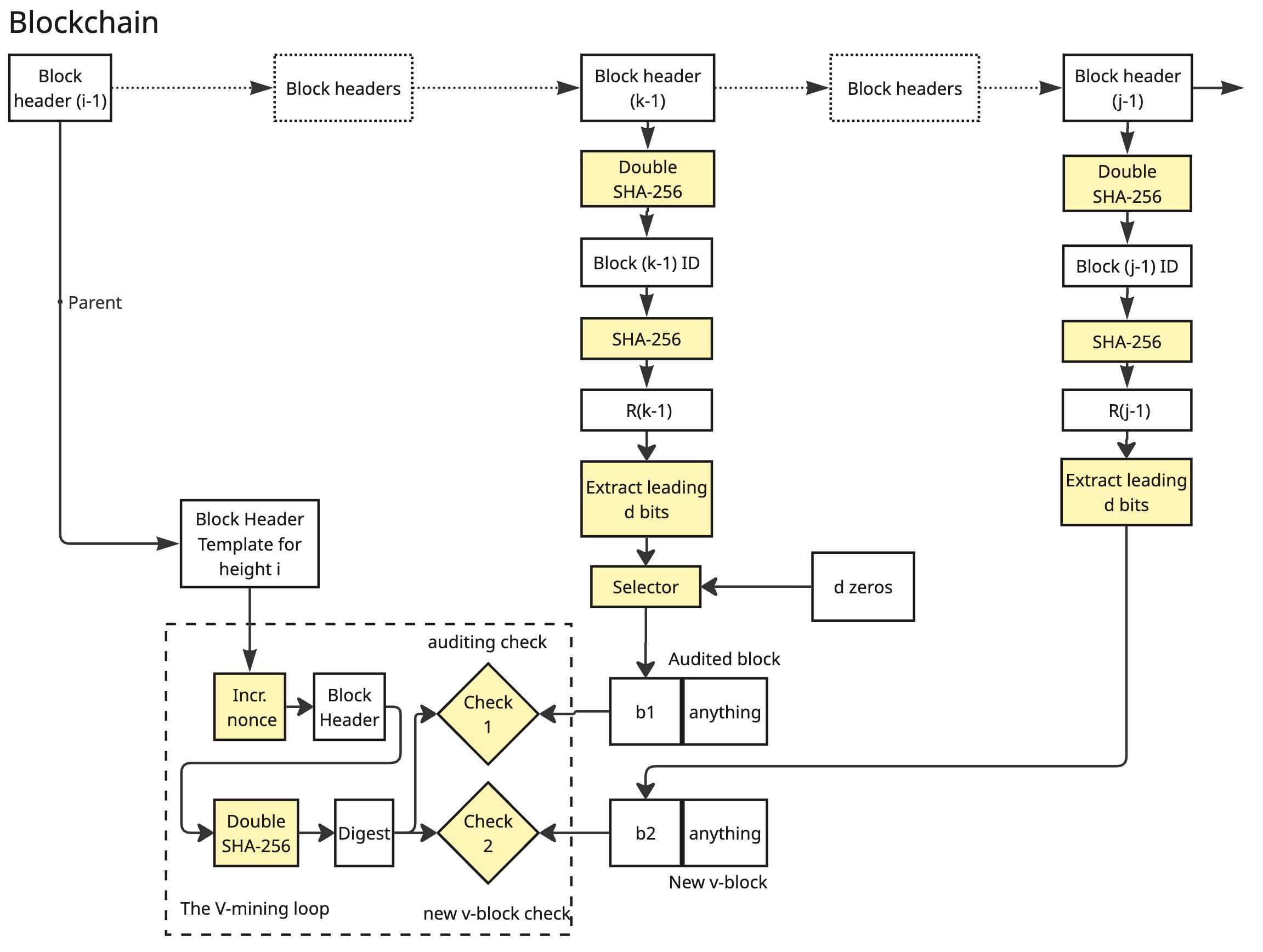}
\caption{Recursive Auditing: audit at height $j$ conducted for an auditing work unit at height $k$, which in turns audits a work unit at height $i$}
\label{fig:V-Mining2}
\end{figure}

The pool could perform audits with smaller portions of the mining power recursively until the potential damage from BWA is negligible. Figure ~\ref{fig:V-Mining2} illustrates the process in which a v-miner audits another v-mining .

\paragraph{Audit cheating.} A miner Mallory doing v-mining may collude with the miner Mallet being audited  and not report Mallet's missing block solutions. This does not pose a high risk since the detection of BWA is probabilistic and the pool server has many opportunities to audit the same miner. For example, the pool server may retain and delay rewards for a predefined number of blocks (i.e. 100 blocks), and the retained reward works as a stake. Even if Mallory can get away with block withholding once, doing it continuously will result in her being discovered and her stake forfeit.
\section{implementation}

In this section, we analyze the requirements and limitations when implemeting APoW on Bitcoin.

\subsection{Compatibility with Bitcoin ASICs}

We believe that our scheme is not compatible with the majority of Bitcoin ASICs.  The minimum Bitcoin difficulty 1 requires checking 32 leading zeros. Requiring 32 zero leading bits for v-mined $X$'s would reduce the RAM cost and make caching a possible alternative to hashing. If mining ASICs allowed checking for only 20 leading zero bits, and nonces for partial matches could be transferred fast enough to the host CPU to be fully checked, then RAM for caching would still be extremely expensive (10M USD per machine) and our v-mining scheme with the addition of a small 20-bit zero-prefix pow would be compatible with those ASICs. However, we have not tested this possibility on actual ASICs.

If pipelined hashing becomes $2^{32}$ times faster/more energy efficient than RAM access, it may be possible to use standard Bitcoin ASICs with Bitcoin difficulty 1. However, as both ASIC speed and RAM density depend on Moore's law, this seems improvable.

The cleanest and safest option is to make a small change in the ASIC hashing circuits, so instead of checking if the leading bits of the hash digest $X$ are zeros, the circuit can also check if the leading bits match a dynamically loaded bit pattern. The bit pattern will consist of the bits $b_2$.

\subsection{Compatibility with Oblivious Shares}

Our method is compatible with oblivious shares. Although not particularly useful, a v-miner can audit a search space of oblivious shares, and a v-miner could submit oblivious shares while auditing. The optimal combination of the two schemes is that a pool can opt to use one or the other. This means that centralized mining pools can use oblivious shares and decentralized mining pools can use \APoW, and both can fairly compete. 

\subsection{Migrating Bitcoin PoW to \APoW}

While switching to a new mining function and disposing all pre-existing ASICs is unfeasible, Bitcoin could hard-fork to enable both methods of mining to co-exist. Mining pools could offer an additional incentive for miners that allow v-mining, and with enough hashrate v-mining, no miner would require to pass KYC checks or amass reputation.

Bitcoin mining hardware has an electrical and economic lifetime of  2-3 years. Therefore, Bitcoin could see most of its mining infrastructure adapted to the new mining function after this period.

\subsection{APoW Without Blockchain Rewards}

\APoW is most effective when the underlying blockchain protocol explicitly rewards the submission of verification blocks (v-blocks), such rewards are not strictly required for the construction to remain useful. Even in the absence of direct compensation for hash power devoted to auditing, a mining pool can still safely assign resources to verification tasks: shares generated during auditing are indistinguishable from ordinary shares and can therefore be accounted for under the pool’s existing payout mechanisms. As long as the expected cost of auditing remains lower than the expected loss induced by a successful block withholding attack, the pool can rationally fund audits without exposing itself to additional avenues of cheating.

\subsection{Defining Auditable Work Units in Mining Pools}
\label{sec:nonce-assignment}

To build the search space $S_i^A$ of the miner $A$, the mining pool must assign a set of work units whose nonce range is large enough so that the set $S_i^A$ is small and can be efficiently communicated to the auditor. Therefore, mining pools must retain some control over the assignment of mining templates and nonce ranges. For every work unit $(G,n_s,n_e)$ sent to a miner, we define the auditable work unit as $(G,n_s,n'_e)$ where $n'_e \leq n_e$, and $n'_e$ is the nonce of the last share submitted by that miner for template $G$.

While workers can propose specific transactions to be included in their templates, certain behaviors that are common in existing Stratum-based protocols must be restricted in order to ensure that the pool can reason unambiguously about which portions of the search space were explored by a given worker.

\paragraph{Disabling timestamp rolling.}
A critical deviation from standard Stratum mining is the prohibition of timestamp (\texttt{nTime}) rolling by workers. If miners are allowed to arbitrarily modify the timestamp field, it becomes infeasible for the pool to determine which nonce spaces correspond to which effective proof-of-work instances, as the hash input changes across timestamps.

Moreover, timestamp rolling enables a strategic block withholding attack in which a miner searches for shares at a given timestamp $t$, and upon finding a share, increments $t$ and continues mining. If a full solution is discovered, the miner can again increment $t$ and resume mining, submitting only shares associated with different timestamps. From the pool’s perspective, the submitted shares appear statistically consistent with honest behavior, while any discovered block solutions are suppressed. To prevent this class of attacks, timestamp rolling must be disabled, and the timestamp must be fixed and assigned by the pool as part of the mining template.

\paragraph{Share frequency and audit granularity.}
The pool selects the share difficulty such that, in expectation, each worker submits at least one share per assigned timestamp value. Auditing is then performed only over nonce interval defined by the starting nonce $n_s$ bounded by the last share submitted by the same worker under the same timestamp. This ensures that the audited region corresponds to a nonce space that was fully explored between observable events, rather than a partially explored suffix of an ongoing search.

\paragraph{Excluding partially explored nonce spaces.}
Auditing must not target nonce values beyond the nonce contained in the worker’s last submitted share for a given template. A worker may legitimately stop mining before completing the full assigned nonce range—for example, if a new block is found by another miner and the pool distributes a new template. In such cases, a valid solution may still exist in the unexplored remainder of the nonce space. Auditing this partially explored region would risk falsely penalizing an honest miner. Accordingly, audits are restricted to nonce intervals that are provably delimited by submitted shares.

\paragraph{Residual withholding and probabilistic deterrence.}
A miner can still attempt to cheat in a limited fashion. Specifically, if the miner finds a full block solution before a share is discovered, the miner may withhold the solution until either a share is found or the template expires. If no share is found during the template’s lifetime, the miner escapes detection for that interval.

However, this strategy is probabilistically constrained. In practice, pools refresh block templates frequently (e.g., every 30 seconds), and share difficulty is typically set so that an honest worker submits shares at a much higher rate (e.g., one share per second on average). Under these parameters, the probability that a solution is found before the share, or that no share is found during the template’s lifetime, is very low.

\section{Summary}

This paper introduces a novel proof-of-work construction designed to deter block withholding attacks (BWAs) in mining pools by enabling auditable reuse of prior nonce-space exploration. Unlike conventional hash-based PoW schemes, where unsuccessful mining effort leaves no verifiable trace, the proposed design allows miners to probabilistically attest—while mining new blocks or shares—that they have exhaustively searched specified regions of the nonce space in previous rounds. This auditability enables mining pools or decentralized pool infrastructures to retroactively verify claimed effort and to detect withholding behavior without relying on trusted hardware, or inherently centralized information-hiding protocols.

We analyze prior work on BWA deterrence, with particular focus on Rosenfeld’s oblivious shares proposal, and identify key limitations related to pool centralization, and latency in block propagation. In contrast, our approach avoids reliance on pool-held secrets or miner ignorance of block validity, and instead supports decentralized pool architectures by allowing independent verification of past mining effort using publicly verifiable information.

We formalize the construction, define its mining and verification processes, and describe how miners can be assigned to probabilistically re-scan prior nonce spaces (“v-mining”) as part of ongoing mining activity. We analyze potential attacks, including proof-of-work reuse, result caching, and collusion during verification, and show that these attacks either offer negligible advantage or can be mitigated through probabilistic auditing and stake-based incentives.

Finally, we discuss deployment considerations, including the fact that the proposed PoW requires a consensus-level change, affects light clients, and is not directly compatible with existing Bitcoin ASICs. While these factors imply substantial adoption friction for existing blockchains, the construction illustrates a new design space for proof-of-work functions that embed auditability as a first-class property, enabling stronger resistance to block withholding and improved support for decentralized mining pools.

\appendix

\section{Zero-Knowledge Proof of Proof-of-Work with Public Payment Address}
\label{app:znp}

\subsection{Preliminaries}

Let
\[
H : \{0,1\}^* \rightarrow \{0,1\}^{\lambda}
\]
be a cryptographic hash function with output length $\lambda$, modeled as a random oracle.
Let $d \leq \lambda$ be a public difficulty parameter.

Let
\[
b \in \{0,1\}^d
\]
be a public bit pattern specifying the required hash prefix, and let
\[
G \in \{0,1\}^g
\]
be a public template fixing certain bit positions of a message.
We denote by $\mathsf{Prefix}_d(z)$ the first $d$ bits of a bitstring $z$.

Let $\mathsf{Addr}$ denote the address space of a public ledger (e.g., Bitcoin addresses).

---

\subsection{Message Structure}

Let $\mathcal{M}(G) \subseteq \{0,1\}^*$ be the set of messages consistent with the public template $G$.
Messages are constructed as
\[
M := \mathsf{Embed}(G, n),
\]
where:
\begin{itemize}
  \item $n \in \{0,1\}^k$ is a private nonce,
  \item $\mathsf{Embed}$ is a deterministic and publicly known embedding function that fills the unfixed positions of $G$ using $n$.
\end{itemize}

Thus, $M \in \mathcal{M}(G)$ if and only if $M$ agrees with $G$ on all fixed template positions.

---

\subsection{Relation Definition}

We define the NP relation $R_{\mathsf{PoW}}$ as follows:
\[
R_{\mathsf{PoW}}((G,b); M) = 1
\iff
\begin{cases}
M \in \mathcal{M}(G), \\
\mathsf{Prefix}_d(H(M)) = b.
\end{cases}
\]

Notably, the relation $R_{\mathsf{PoW}}$ is independent of the payment address.

---

\subsection{SNARK Formulation}

Let
\[
\mathsf{SNARK} = (\mathsf{Setup}, \mathsf{Prove}, \mathsf{Verify})
\]
be a non-interactive zero-knowledge succinct argument for NP.

\paragraph{Public inputs.}
\[
x := (G, b, \mathsf{addr})
\]
where $\mathsf{addr} \in \mathsf{Addr}$ is a public payment address on the ledger.

\paragraph{Private witness.}
\[
w := M
\]

\paragraph{Prover.}
\[
\pi \leftarrow \mathsf{Prove}(x, w)
\]

\paragraph{Verifier.}
The verifier accepts if and only if
\[
\mathsf{Verify}(x, \pi) = 1.
\]

---

\subsection{Security Properties}

\paragraph{Completeness.}
If the prover knows a message $M$ such that
$R_{\mathsf{PoW}}((G,b); M)=1$,
then for any payment address $\mathsf{addr}$, the verifier accepts with overwhelming probability.

\paragraph{Soundness.}
No probabilistic polynomial-time adversary can produce a valid proof $\pi$ for public input $(G,b,\mathsf{addr})$ without knowing a message $M$ such that
\[
M \in \mathcal{M}(G)
\quad \wedge \quad
\mathsf{Prefix}_d(H(M)) = b.
\]

\paragraph{Zero-Knowledge.}
The proof $\pi$ reveals no information about the message $M$ (and thus about the nonce $n$), beyond the fact that a valid proof-of-work solution exists.

---

\subsection{Reward Attribution}

Since the payment address $\mathsf{addr}$ is included as a public input, any verifier accepting a proof $\pi$ for $(G,b,\mathsf{addr})$ can attribute the corresponding proof-of-work to that address and issue rewards on the underlying ledger, while the PoW solution itself remains hidden.

\section{Unsealed V-Blocks}
\label{app:unsealed-vblocks}

Creating empty unsealed v-blocks has the benefit that simplifies the security analysis and minimally affects the properties of the Nakamoto consensus protocol. 

There are different protocols design choices for how to reference the empty v-block in the blockchain. We present 3 methods: 

\begin{enumerate}
    \item \textbf{Scheme 2a.} The v-block header is included as in the block header chain and processed as a block, but only executes a coinbase transaction that pays the miner.
    
    \item \textbf{Scheme 2b.} The v-block header is not part of the block header chain. Instead, it is referenced by one future block header up to a maximum distance, similar to GHOST\cite{sompolinsky2015ghost} uncles. Nevertheless, the v-block propagates independently over the p2p network. Also, depending on the design, the v-block header may or may not contribute to the cumulative difficulty of the chain. GHOST stipulates that uncle blocks do contribute. Independently of v-blocks work contribution, the protocol must mint a reward for the v-block header to create incentives for v-mining.  

    \item \textbf{Scheme 2c.} The v-block header is not part of the blockchain header chain, but is included in the blockchain inside a normal transaction $T$ with a special payload (i.e., OP\_RETURN data in Bitcoin, or calldata in EVM-like chains). The v-block does not contribute to the cumulative work of the chain, but the protocol mints a subsidy $S$ to reward the v-miner. The transaction $T$ has an output with value $S$, which is considered as a virtual input of the transaction, same as the coinbase. 
\end{enumerate}

For Scheme 2c, we set a limit on the number of v-blocks that can be created for a specific height (i.e, 2 v-blocks per height) to prevent miners from creating only v-blocks. For scheme 2b, GHOST already limits the number of uncles.

Mining v-blocks should be almost as profitable as mining normal blocks to avoid penalizing pools that use them to perform audits. However, it should not be equally profitable. If so, the mining pools could engage only in v-mining, reducing the available transaction volume on the blockchain and preventing other pools from auditing. 
In Bitcoin, v-mining would be discouraged as unsealed v-blocks cannot earn transaction fees.



We now present how a v-block is broadcast and validated for the Scheme 2c. 

\paragraph{Broadcast and validation.}

Let $j$ be the v-mining height auditing a work unit at height $i$. After finding the PoW, the v-miner creates a transaction $T$ containing ($G_i,n$) and indicating that it contains a v-mined block header. With this information, the message $M$ can be rebuilt. Let $t$ be the block height of the block containing $T$ being processed, or, if $T$ is being forwarded on the mempool, the height of the block following the tip of the canonical chain. The transaction is only valid if $M$ corresponds to a valid v-block header. A v-mined block is valid if it satisfies the following conditions:

\begin{enumerate}
    \item $\mathsf{version}$ must be valid
    \item $\mathsf{parent}$ block must exist in the canonical chain at height $(j-1)$ and $ 0 < t - j \leq D_I$, for a maximum inclusion depth $D_I$.
    \item $\mathsf{height}$ must be $j$.
    \item $\mathsf{time}$ is ignored.
    \item $\mathsf{difficulty}$ must respect the difficulty requirements for block $i$.
    \item The PoW of $M$ must be valid according to the declared difficulty.
    \item $\mathsf{poolAddress}$ is a valid payment address for the ledger.
\end{enumerate}

When the transaction is included in a block and processed, it pays the current block subsidy to a predetermined address, although without additional transaction fees, since no transaction is specified in the v-block header.

To implement this scheme in Bitcoin, the $\mathsf{height}$ and $\mathsf{poolAddress}$ must be stored in the first input scriptSig field of the coinbase transaction, so the coinbase transaction and an associated Merkle path must also be included in $T$ along the v-block header, making it larger and potentially more expensive. To avoid oversized transactions, the field $\mathsf{txroot}$ could be replaced by a new field $\mathsf{dataRoot}$  that represents the root of a Merkle tree of data fields, where $\mathsf{height}$, $\mathsf{poolAddress}$ and $\mathsf{txroot}$ are some of its terminal elements.

\bibliographystyle{alpha}

\end{document}